\documentclass[12pt]{article}
\usepackage{epsf}
\def\be{\begin{equation}}
\def\ee{\end{equation}}
\def\bea{\begin{eqnarray}}
\def\eea{\end{eqnarray}}

\begin{document}
\begin{titlepage}
\begin{center}
{\Large \bf William I. Fine Theoretical Physics Institute \\
University of Minnesota \\}
\end{center}
\vspace{0.2in}
\begin{flushright}
FTPI-MINN-17/24 \\
UMN-TH-3707/17 \\
December 2017 \\
\end{flushright}
\vspace{0.3in}
\begin{center}
{\Large \bf Equality of $e^+e^-$ production amplitudes for  scalar-vector and pseudoscalar-axial heavy meson-antimeson pairs.
\\}
\vspace{0.2in}
{\bf  M.B. Voloshin  \\ }
William I. Fine Theoretical Physics Institute, University of
Minnesota,\\ Minneapolis, MN 55455, USA \\
School of Physics and Astronomy, University of Minnesota, Minneapolis, MN 55455, USA \\ and \\
Institute of Theoretical and Experimental Physics, Moscow, 117218, Russia
\\[0.2in]

\end{center}

\vspace{0.2in}

\begin{abstract}
The production of heavy meson-antimeson pairs of the type $S\,V$ and $P \, A$ in $e^+e^-$ annihilation is considered, with $P$ and $V$ being the ground-state $J^P=0^-$ and $J^P=1^-$ (anti)mesons from the $(1/2)^-$ doublet, and $S$ and $A$ standing for the excited $J^P=0^+$ and $J^P=1^+$ (anti)mesons from the $(1/2)^+$ doublet. It is argued that the production amplitudes in these two channels should be equal up to a higher (than one) order in the heavy quark mass ($\Lambda_{QCD}/M_Q$) expansion,  $A(e^+e^- \to S \bar V) = A(e^+e^- \to A \bar P)$, including both the $S$-wave and the $D$-wave amplitudes. Given that the $S\,V$ and $P \, A$  thresholds are extremely close, the production cross section in both channels should be the same to a high degree of accuracy.  In practice this behavior can be studied for the processes $e^+e^- \to D_{s0}(2317) \bar D_s^* +$ c.c. and  $e^+e^- \to D_{s1}(2460) \bar D_s +$ c.c. in the charm sector  and $e^+e^- \to B_{s0} \bar B_s^* +$ c.c. and $e^+e^- \to B_{s1} \bar B_s +$ c.c. in the $B$ sector.
\end{abstract}
\end{titlepage}

Recently the BESIII experiment has reported~\cite{bes} an observation and analysis of the final state $D_{s0}(2317) \bar D_s^* +$ c.c. produced in the $e^+e^-$ annihilation at c.m. energy up to 4.6\,GeV. In addition to other results of the study, this is in fact the first experimental evidence of a mixed $(1/2)^+ + (1/2)^-$ heavy meson pair production. The final states of this type are singled out among other combinations of heavy meson-antimeson pairs by that only for these a direct $S$-wave production in the $e^+e^-$ annihilation is allowed, and the observed angular distribution~\cite{bes} is consistent with the $S$-wave. Another well known~\cite{pdg} peculiarity of the charmed-strange $(1/2)^+$ meson doublet is that the mass splitting between its heavier and lighter components $D_{s1}(2460) - D_{s0}(2317)$, $141.8 \pm 0.8$\,MeV, is almost exactly the same as in the ground-state $(1/2)^-$ doublet $D_s^* - D_s$: $143.8 \pm 0.4$\,MeV. For this reason the thresholds in the channels  $D_{s0}(2317) \bar D_s^* +$ c.c. and $D_{s1}(2460) \bar D_s +$ c.c. are split by only about 2\,MeV and are thus almost degenerate, and a very similar (approximate) degeneracy is also expected for the bottom-strange mesons~\cite{beh,kl,gscpz}. [For the non strange heavy mesons such degeneracy is rather moot, since the excited $(1/2)^+$ mesons have large width (up to 250 - 400 MeV) due to $S$-wave decay to a pion and their $(1/2)^-$ counterpart, e.g. $D_0^*(2400) \to D \, \pi$.]  Furthermore, it has been argued~\cite{mv17} on the basis of the heavy quark spin symmetry (HQSS) that in the limit of exact degeneracy only one coherent  combination of the two channels is produced in $e^+e^-$ annihilation: $(V_i \bar S - \bar V_i S + A_i \bar P - \bar A_i P)$, where $P$ and $V$ stand for the pseudoscalar and vector mesons in the ground state doublet and $S$ and $A$ denote the scalar and axial mesons in the excited doublet. The coherence between the channels can be important in near-threshold `molecular' resonances, if such states exist, but the predictions from HQSS for the relative production of various meson-antimeson channels are usually strongly violated, as is known since long ago~\cite{drgg}, at the energy distance from the relevant thresholds that is comparable with the spin splittings between the meson masses~\cite{mv12}. The purpose of the present paper is to point out that the equality of the yield in the channels $S \, V$ and $P \, A$ is protected, in addition to the HQSS, by the chiral symmetry and parity conservation. The additional constraint arises in a scheme~\cite{beh} where the observed equality of the mass splittings in the doublets is understood by considering the mesons in the $(1/2)^+$ and $(1/2)^-$ doublets as `parity doubles'. 

Namely, if the mesons are described by the heavy spin multiplets $(S,\, A_\mu)$ and $(P,\, V_\mu)$ in the matrix form:
\be
{\cal H} = {1 \over 2} \, (i \gamma^5 P + \gamma^\mu V_\mu) (1+ \gamma \cdot v),~~~~{\cal H'} = {1 \over 2} \,  (i \gamma^5 S + \gamma^\mu A_\mu) (1+ \gamma \cdot v)
\label{defh}
\ee
with $v$ being the 4-velocity, the combinations 
\be
{\cal H}_L ={1 \over \sqrt{2}} ({\cal H} - i {\cal H'}),~~~~{\cal H}_R ={1 \over \sqrt{2}} ({\cal H} + i {\cal H'})
\label{hlr}
\ee
transform as $(3,1)$ and $(1,3)$ representations under the chiral $SU(3)_L \times SU(3)_R$ symmetry, where the flavor $SU(3)$ indices for the heavy mesons are understood.
Clearly, the latter property implies that the spinless and spin-one mesons of the appropriate chirality:
$H_L = (P- iS)/\sqrt{2}$, $H_R=(P+ iS)/\sqrt{2}$, $H^\mu_L= (V^\mu - i A^\mu)/\sqrt{2}$ and $H^\mu_R= (V^\mu + i A^\mu)/\sqrt{2}$, have definite transformation properties under the chiral symmetry.

The meson-antimeson pairs under the present consideration can be produced in $e^+e^-$ annihilation in the $S$- and $D$-waves. The data~\cite{bes} indicate that the production in the channel $D_{s0}(2317) \bar D_s^* +$ c.c. at 4.6\,GeV (i.e. approximately 170\,MeV above the threshold) is mostly contributed by an $S$-wave. However there can be some contribution from  a $D$-wave whose relative significance can grow at higher energy. Thus one should generally consider both terms in the amplitude of the production of the meson pairs. The heavy meson pairs are produced by the electromagnetic current $J_\mu$ of the corresponding heavy quark (charmed or bottom) which is a singlet under the chiral $SU(3)_L \times SU(3)_R$ symmetry, as well as being parity negative. These constraints result in a unique expression for the amplitude that can be written in terms of the wave functions of the chiral combinations of the mesons. In the center of mass of the production process , where the current has only spatial components: $\vec J$, this expression has the form
\bea
 && J_\ell= i \left [ C_S \delta_{\ell j} + C_D \left ( p_\ell p_j - {1 \over 3} \delta_{\ell j} {\vec p}^{\,2} \right ) \right ] \, \left [ \left ( \bar H_L { H}_{L j}- \bar H_R { H}_{R j} \right ) - \left ( \bar H_{L j} H_L -  \bar H_{R j} H_R \right ) \right ] = \nonumber \\
&&\left [ C_S \delta_{\ell j} + C_D \left ( p_\ell p_j - {1 \over 3} \delta_{\ell j} {\vec p}^{\,2} \right ) \right ] \, \left ( V_j \bar S - \bar V_j S + A_j \bar P - \bar A_j P \right )
~,
\label{curr}
\eea
where ${\vec p}$ is the vector of the c.m. momentum and $C_S$ and $C_D$ are scalar $S$- and $D$-wave amplitude factors that are (generally complex) functions of the c.m. energy. Clearly, the presence in Eq.(\ref{curr}) of the $\bar L \times L$ and $\bar R \times R$ products results from the requirement for the current to be a chiral singlet, the relative coefficient (with the minus sign) is due to the parity, and the relative sign for the charge conjugate channels is due to the $C$ conjugation property of the current. It is also clear that those are the first two requirements that set the relative coefficient of the ${\bar S}V$ and $\bar P A$ production amplitudes determining the relative yield in the two discussed channels not related by the charge conjugation. It should be noted that the expression in Eq.(\ref{curr}) does not rely on any assumptions about the full  form of the current written in terms of full heavy symmetry multiplets from Eq.(\ref{defh}). The latter assumptions would relate the amplitude in the considered channels to that in the heavier channel $A \bar V - \bar A V$ with any such relations being fully dependent on HQSS and generally affected by a significant violation in the threshold energy region.  Thus it should be expected that the relations between the production amplitudes, following from Eq.(\ref{curr}): 
\bea
&&A(e^+e^- \to D_{s0}(2317) \bar D_s^*) = A(e^+e^- \to D_s \bar D_{s1} (2460) )~, \nonumber \\
&&A(e^+e^- \to B_{s0} \bar B_s^*) = A(e^+e^- \to B_s \bar B_{s1}  )~,
\label{arel}
\eea
should hold with accuracy beyond the first order in the HQSS breaking. In particular these relations do not receive a first order correction due to the term in the effective chiral-symmetric Lagrangian describing the HQSS violating mass splittings in both doublets~\cite{beh}
\be
{\cal L}_{hyperfine} = {\kappa \, \Lambda_{QCD}^2 \over 12 M_Q^2} {\rm Tr} \left ( \bar {\cal H}_L \sigma_{\mu \nu} {\cal H}_L \sigma^{\mu \nu} + \bar {\cal H}_R \sigma_{\mu \nu} {\cal H}_R \sigma^{\mu \nu} \right ) ~,
\label{dm}
\ee
where $\kappa$ is a coefficient of order one.

It is clear however that the HQSS violating effects do modify in a higher order the degeneracy between the two discussed channels. One such effect is a purely kinematical one due to the mass splitting in the doublets. It lifts the degeneracy both at the level of the amplitudes and at the level of the cross section due to a (slightly) different c.m. momentum $p$ in the two channels. Indeed, assuming that the mass splitting $\delta$ is the same in the two doublets: $M_V = M_P + \delta$, $M_A=M_S+\delta$, and using the notation $\Delta = M_S - M_P$, one readily finds at $E_{c.m.}^2 = s$ the ratio of the c.m. momenta in the two channels
\be
{p_{c.m.} (SV) \over p_{c.m.}(AP)} = \left [ {s - (M_S-M_V)^2 \over s- (M_A-M_P)^2} \right ]^{1/2} = \left [ {s - (\Delta - \delta)^2 \over s- (\Delta+ \delta)^2} \right ]^{1/2} \approx 1 + {2 \Delta \, \delta \over s} \approx 1+ O \left ( {\Lambda_{QCD}^3 \over  M_Q^3} \right )~,
\label{ratp}
\ee
where in the last transition it is taken into account that $\Delta \sim \Lambda_{QCD}$, $\delta \sim \Lambda_{QCD}^2/M_Q$, and $s \sim 4 M_Q^2$ (or larger).

A potentially larger than the purely kinematical effect may arise in the order $\Lambda_{QCD}^2/M_Q^2$ from the mixing of the axial components of $(1/2)^+$ and $(3/2)^+$ heavy meson doublets (a discussion of such mixing for non strange $B$ mesons can be found e.g. in Ref.~\cite{bv}). The HQSS violating mixing, whose amplitude is generally of order $\Lambda_{QCD}/M_Q$, would enter quadratically in the relations (\ref{arel}) between the production amplitudes. Indeed, if $\theta \sim \Lambda_{QCD}/M_Q$ is the mixing angle, the coefficient of the axial $A$ in Eq.(\ref{curr}) is modified by the factor $\cos \theta \sim 1- O(\Lambda_{QCD}^2/M_Q^2)$. It is not clear at present how to estimate numerically a deviation from the equalities (\ref{arel}). However, one may notice that the same mixing slightly shifts down (also in the second order) the mass of the axial meson in the $(1/2)^+$ doublet. Thus, if the (approximately) 2\,MeV difference between the mass splittings in the $(1/2)^+$ and $(1/2)^-$ charmed doublets is attributed to this effect, it significance, in the scale of $\Lambda_{QCD}$ can be estimated as being of order one percent. It is then reasonable to expect a modification of the relations (\ref{arel}) by about the same amount in the charm sector. 

As discussed, the equalities (\ref{arel}) between the production amplitudes rely on the chiral symmetry scheme~\cite{beh}, which symmetry is also only approximate and is violated by the light quark masses. This violation is especially significant for the strange heavy mesons due to a larger strange quark mass. However the effect of this mass in the first order describes the flavor $SU(3)$ breaking differences between the non strange and strange heavy mesons, and apparently is absent (in this order) from the relations for the `parity doubles' of the same flavor. In particular, there is no significant breaking of the discussed degeneracy of the mass splittings in the $(1/2)^+$ and $(1/2)^-$ charmed strange meson doublets.

The production amplitudes in Eq.(\ref{arel}) are necessarily complex, e.g. due to a rescattering between channels with and without the hidden strangeness. It should thus be emphasized that the claimed here equalities between them imply that their absolute values should be the same (equal cross sections) and also their complex phases are the same. It can be noticed that both discussed channels end up in the same final state after strong and electromagnetic decays of excited heavy mesons, e.g. the meson pair
$D_{s0}(2317) \bar D_s^*$ end up in $D_s \bar D_s \pi^0 \gamma$ after the decays $D_{s0}(2317) \to D_s \pi^0$ and $\bar D_s^* \to \bar D_s \gamma$, while the pair $D_s \bar D_{s1} (2460)$ results in the same set of particles after $\bar D_{s1} (2460) \to \bar D_s^* \pi^0 \to \bar D_s \gamma \pi^0$. Thus a study of the relative phase of the production amplitudes by interference effects is possible in principle, however in practice the interference effects are greatly reduced by a very small region of the $D_s \bar D_s \pi^0 \gamma$ phase space where the products from the two process chains kinematically overlap.  

In summary. It is argued that within the chiral symmetry scheme, previously developed~\cite{beh} for description of the spectra of the ground-state and excited heavy mesons, the amplitudes of production in $e^+e^-$ annihilation of mixed scalar-vector and axial-pseudoscalar pairs of heavy mesons are equal as described by Eq.(\ref{arel}). In particular this implies the equality of the cross sections: $\sigma[e^+e^- \to D_{s0}(2317) \bar D_s^* + {\rm c.c.}] = \sigma[e^+e^- \to D_s \bar D_{s1} (2460) + {\rm c.c.}]$. The relation should hold with accuracy better than the first order in HQSS breaking and likely better than the first order in the chiral symmetry breaking. An estimate of a specific numerical accuracy of the relation is not quite clear at present, but it can be as low as  `in the ballpark' of a few percent. An observation of the former production process is recently reported~\cite{bes} by BESIII. Thus a test of the discussed here equality appears to be within the reach of present experiments.

This work is supported in part by U.S. Department of Energy Grant No.\ DE-SC0011842.


\begin{thebibliography}{99}
\bibitem{bes} 
  M.~Ablikim {\it et al.} [BESIII Collaboration],
  arXiv:1711.08293 [hep-ex].
	
\bibitem{pdg} 
  C.~Patrignani {\it et al.} [Particle Data Group],
  Chin.\ Phys.\ C {\bf 40}, no. 10, 100001 (2016).
  doi:10.1088/1674-1137/40/10/100001; \ 2017 on line update
	
\bibitem{beh}
W.~A.~Bardeen, E.~J.~Eichten and C.~T.~Hill,
  Phys.\ Rev.\ D {\bf 68} (2003) 054024 \ 
  [hep-ph/0305049].

\bibitem{kl}
E.~E.~Kolomeitsev and M.~F.~M.~Lutz,
  Phys.\ Lett.\ B {\bf 582} (2004) 39 \ 
  [hep-ph/0307133].

\bibitem{gscpz}
F.~K.~Guo, P.~N.~Shen, H.~C.~Chiang, R.~G.~Ping and B.~S.~Zou,
  Phys.\ Lett.\ B {\bf 641} (2006) 278 \ 
  [hep-ph/0603072].
	
\bibitem{mv17} 
  M.~B.~Voloshin,
  Phys.\ Rev.\ D {\bf 95}, no. 5, 054017 (2017)
  doi:10.1103/PhysRevD.95.054017
  [arXiv:1701.03064 [hep-ph]].
	
\bibitem{drgg} 
  A.~De Rujula, H.~Georgi and S.~L.~Glashow,
  Phys.\ Rev.\ Lett.\  {\bf 38}, 317 (1977).
  doi:10.1103/PhysRevLett.38.317
	
\bibitem{mv12} 
  M.~B.~Voloshin,
  Phys.\ Rev.\ D {\bf 95}, no. 5, 054017 (2017)
  doi:10.1103/PhysRevD.95.054017
  [arXiv:1701.03064 [hep-ph]].
\bibitem{bv} 
  A.~E.~Bondar and M.~B.~Voloshin,
  Phys.\ Rev.\ D {\bf 93}, no. 9, 094008 (2016)
  doi:10.1103/PhysRevD.93.094008
  [arXiv:1603.08436 [hep-ph]].	


\end{thebibliography}
\end{document}